\documentclass[12pt,preprint]{aastex}
\usepackage{amssymb}
\usepackage{txfonts}

\begin{document}

\shorttitle{Cool matter in solar corona} \shortauthors{Qu et al.}

\title{Cool matter distribution in inner solar corona from 2023 total
solar eclipse observation}

\author{Z.Q. Qu$^{1,2}$, H. Su$^{3}$, Y. Liang$^{4}$, Z.Xu$^{1}$, R.Y. Zhou$^{5}$\\
1. Yunnan Astronomical Observatory, CAS, Tianwentai road, Guandu district, Kunming, Yunnan 650216, China\\
2. School of Astronomy and Space Sciences, University of Chinese
Academy of Sciences, Yanxihu, Huairou District, Beijing,
China\\
3. Yunnan Amateur Astronomers Association, No.2506 Jida Square,
Panlong district, Kunming, Yunnan, China\\
4. Shanghai Astronomical Observatory, CAS, Xuhui District,
Shanghai, China\\
5. Department of Physics, Chinese University of Hong Kong, Shatin,
New Territories, Hong Kong SAR, China}

\begin{abstract}
Solar corona has been judged to consist of free electrons and highly
ionized ions with extremely high temperature as a widely accepted
knowledge. This view is changed by our eclipse observations.
Distributions of cool matter represented by neutral iron atoms in
hot inner solar corona are presented via derived global maps of
solar Fraunhofer(F-) and Emission(E-) coronae, compared with those
of continuum(Kontinuierlich, K-) corona formed by free electrons.
The maps are obtained from simultaneous observations of dual
filtering bands centered respectively at 659.4nm and 660.1nm,
performed from twin telescopes during the total solar eclipse on
April 20, 2023 at Com town of East Timor, assisted for judgement via
spectral images obtained by a portable spectrograph. They show
respectively presences of these neutral iron atoms yielding 659.3nm
and 659.4nm lines in both the quiet sun and active regions. The
distribution of the cool matter in form of line depression forms an
inner F-corona, different from that of the cool matter in form of
line enhancement. Both the distributions show a crucial difference
from that of the free electrons represented by the K-corona map. It
is also found that intensities of the F-corona and the E-corona
induced by these neutral atoms are only small fractions of the
K-corona, and the diffusion can be seen clearly in all these maps.
They uncover also that the coronal heating resources do not
distribute pervasively but likely form a thermodynamic griddle where
minor photospheric neutral atoms can escape from the heating into
the corona globally.
\end{abstract}

{Key Words}\hspace{1cm} Sun:corona; Sun: Fraunhofer corona

\section{Introduction}
Solar corona has been being observed during total solar eclipses
since ancient times with human naked eyes. Giovanni Cassini coined
the name after his observation of 1706 solar eclipse. However, it
had been naturally regarded to be cooler than its underneath
atmospheric layers such as solar photosphere with an average
temperature of several thousand degrees since it is farther away
from the solar core, the only heat resource recognized then.
Temperature of the corona was first reliably diagnosed and found to
be hotter than one million Kelvin since Hardness and Young recorded
the green coronal line in the corona during 1869 solar eclipse, and
Grotrian (Grotrian,1939) in 1939 and later Edlen(Edlen,1943) in 1943
realized that this spectral line is caused by transition from upper
energy levels in thirteen times ionized iron ions, whose metastable
presence necessary for yielding observable radiation needs an
environment of more than one million degrees in the atmosphere.
Since then, the opposite view has been accepted that the quiet
corona unconfined by the magnetic fields or in open field line
region, was regarded to be consisted of completely highly ionized
ions and free electrons. This forms the basis of the solar coronal
physics(Aschwanden,2015). On the other hand, beyond a height of
three solar radii above solar limb, dust was found scattering the
Fraunhofer lines originally yielded in the photosphere(Morgan et
al., 2007; Stenborg et al., 2021). It is called F-corona. And
according to space observations, the dust is distributed in an
elliptic space around the sun (Burtovoi et al., 2022;
 Lamy et al., 2022). However, conscious detection for presence of a cool
component has not yet been explored in the inner solar corona, where
the dust should be sublimated (Anderson, 1924; Russell, 1929),
probably due to its potential confliction with the aforementioned
view.

In astrophysics, temperature of a mass of electromagnetically
structured particles can be diagnosed from the specific spectral
lines that are produced with a favorite temperature, causing the
maximum occupation number of the involved energy levels of the
transition, and thus giving the birth to the most strong line
intensity. In detail, this kind of $'formation'$ temperature can be
roughly estimated from the excitation potentials of these levels
excited from the ground ones. For instance, for our used neutral
iron line at 659.3nm and 659.4nm, the excitation energies of their
lower levels are respectively 2.73eV and 2.43eV, and those of their
upper levels are respectively 4.61eV and 4.31eV. As a contrast, the
excitation and ionization energy needed for producing the green
coronal line FeXIV530.3nm is greater than 394.6eV(from NIST,
$https://physics.nist.gov$ $/PhysRefData/ASD/lines$
$\_$$form.html$). Therefore, the environment in which the neutral
atoms yielding these two lines can survive is two order colder than
that producing the green coronal line, provided that the excitation
and ionization energy $E$ is supplied by the collisions of thermal
random motions. That is, $E=3/2kT$ may be approximately applied,
where $T$ means the temperature and $k$ the Boltzmann constant.
Accordingly, the specific spectral lines can be temperature proxies
of mass of these atoms or ions.

It was noteworthy that our spectropolarimetry of upper solar
atmosphere during 2013 Gabon eclipse led to the finding of the
neutral atoms in the inner solar corona(Qu et al., 2024). As pointed
out previously, one way to identify the existence of the cool matter
represented by the neutral atoms is to detect the presence of the
specific spectral lines yielded by these atoms and then to exclude
the possibility that the presence is caused by the dust scattering
due to projective effect along the line-of-sight. The first step can
be taken by performance of simultaneous observations of two bands
with the same-width containing respectively the suitable lines and
their adjacent continua, using two same telescopes and detectors. It
is basically understood that if absolute value of difference between
the intensities of the two bands is well above the detection noise,
a definite evidence for the presence of the relevant atoms producing
the specific spectral lines is implied. Furthermore, if the
intensity difference is definitely positively greater than the noise
level, it indicates that there are spectral lines detected above the
continuum by the neutral atoms producing the emission lines via
transitions from the upper energy levels to the lower ones and/or
scattering these emission lines. And if the difference is negative
and its absolute value well above the noise level, it can be judged
that there are the specific lines whose intensities are depressed
due to the neutral atoms yielding the absorption via transitions
from the lower energy levels to the upper ones, and/or scattering
the Fraunhofer lines predominantly formed in the photosphere. The
aforementioned criterion is the reasoning base in the following.

\section{Observations and Results}
It becomes well known that total solar eclipses provide us uniquely
precious chances to get the accurate measurement of the quiet solar
corona on the ground with minimum light pollution from the telluric
atmosphere and instrument during the totality(Pasachoff, 2017). We
performed the observation during a total solar eclipse occurred on
April 20, 2023, at Com town, East Timor. The geographical
coordinates of the site read as E127$^{\circ}$3$^{'}$42$^{''}$
S8$^{\circ}$21$^{'}$38$^{''}$. The totality last about 76 seconds
from 04:21:34UT to 04:22:50UT. Our observational goal is to survey
the simultaneously available the F-corona described by the selected
Fraunhofer line intensity depression and the E-corona formed by the
same specific line enhancement in the inner corona, compared with
the K-corona formed by the continuum intensity source distribution.
The Fraunhofer lines are selected, as stated above, to be two
neutral iron lines, FeI 659.29nm ($3d^{7}(^{2}G)4s\rightarrow
3d^{7}(^{4}F)4p$) and 659.38nm ($3d^{6}4s^{2}\rightarrow
3d^{7}(^{4}F)4p$). They are contained in a 659.4nm-centered
filtering band with 0.3nm bandpass. The images are obtained from the
twin apochromatic refractive telescopes named APO80. Both the
telescopes have the same primary optical parameters (Aperture 80mm
and focal ratio f/6) except two different filters mounted
respectively in front of two same electronic detectors Sony IMX174.
Thus, one telescope provides via a filter(type: Alluxa
7337,659.4-0.3nm) images of the band 659.25-659.55nm containing the
two lines mixed with continuum, and the other genders images of the
continuum via a filter(type: Aulluxa 7340, 660.1-0.3nm) with the
same bandpass width 0.3nm but centered at 660.1nm. These twin
telescopes are installed symmetrically on one dovetail plate
attached to an equatorial mount. Finally, in order to check whether
additional lines are present within the two band thus possibly
distort the intensity difference between the two bands, a portable
spectrograph named Sol$'$Ex with a guiding optics of 8cm aperture
refractive telescope was put into actual observation, whose
observational band covers the two filtering bands as well as the
famous H$\alpha$ line.

Before we carried out the observation, an integrating sphere with a
stable light in our laboratory had been used for intensity
calibration between the two telescopes and test for readout
linearity of the detector. Further intensity calibration between the
two channels was performed in a way that their intensity difference
should be minimized in the dark disk formed by the lunar occultation
during the totality after the data reduction of dark current and
flat-fielding. And since the two kinds of images were recorded at
the different parts of the detector chips and influenced by
seeing-induced image movement, the location correlation registration
is carried out before the further data reduction operation. Finally,
the intensity differences at those points are set to be zero whose
intensities obtained from both the two bands are saturated.

The top two panels in Fig.1 show the sample images of the two bands
obtained respectively from the twin telescopes with the same cadence
and exposure times of 0.1 second for each frame during the total
solar eclipse. Both of these images are respectively superpositions
of thirteen raw frames after the data reduction. The left one is the
filtering image within the band centered at 659.4nm containing the
two lines, with intensity indicated by $I_{c,l}$. And the
K-corona(continuum intensity) image is plotted in the right panel
acquired within the band centered at 660.1nm, with its intensity
indicated with $I_{c}$. It is intuitively very hard to distinguish
one image from the other. However, quantitative comparison will
reveal their delicate difference and this is just our observational
goal. It is notable that the regions with strong radiation are
regarded to be active ones.

\begin{figure}
\flushleft
\includegraphics[width=18.0cm,height=5.6cm]{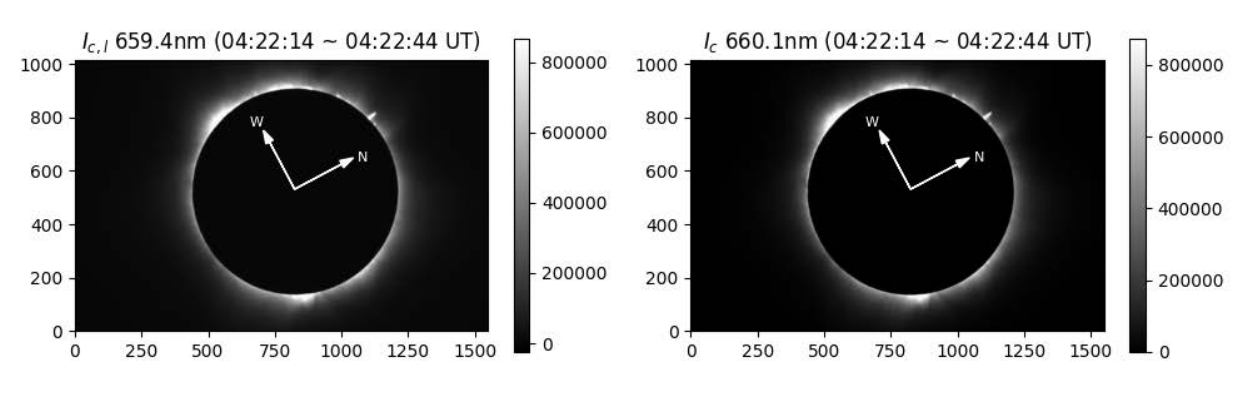}
\centering
\includegraphics[width=13.8cm,height=7.2cm]{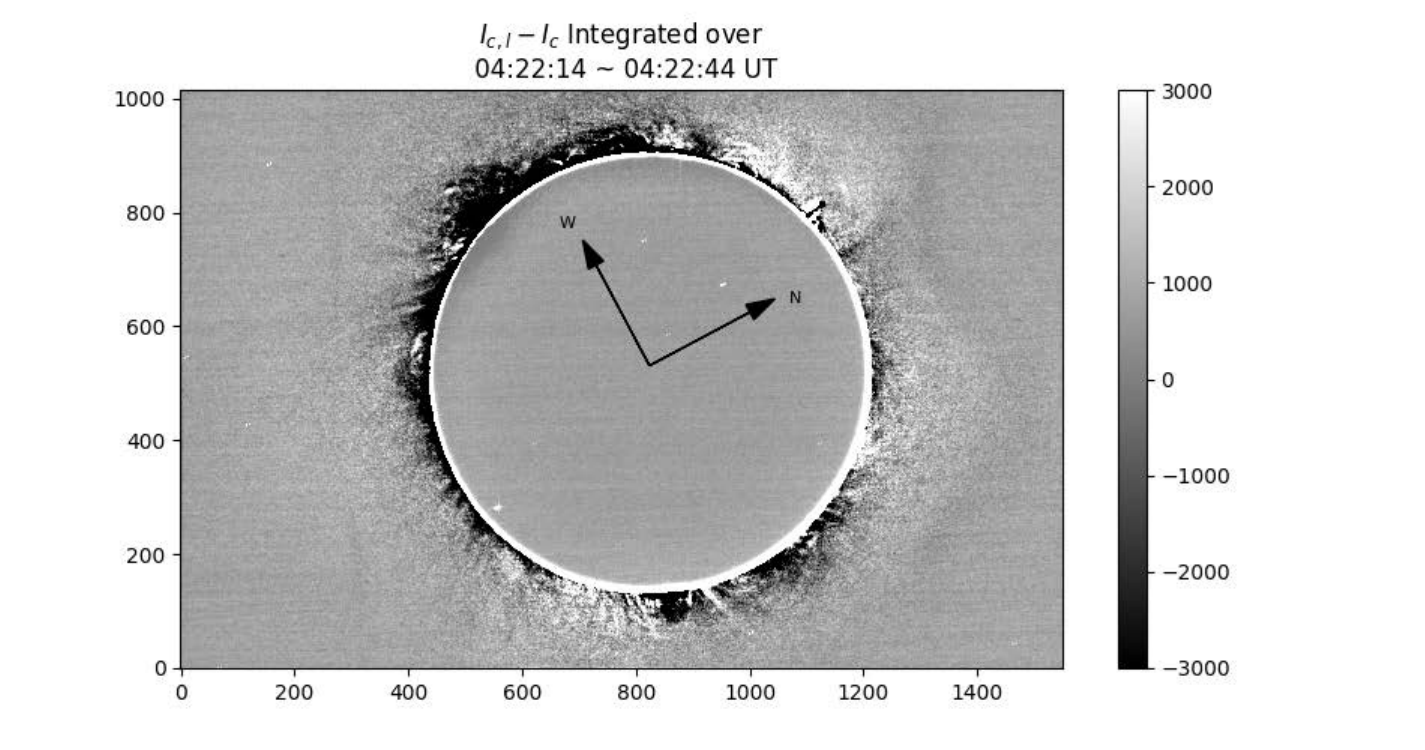}
\flushleft
\includegraphics[width=17.8cm,height=6.0cm]{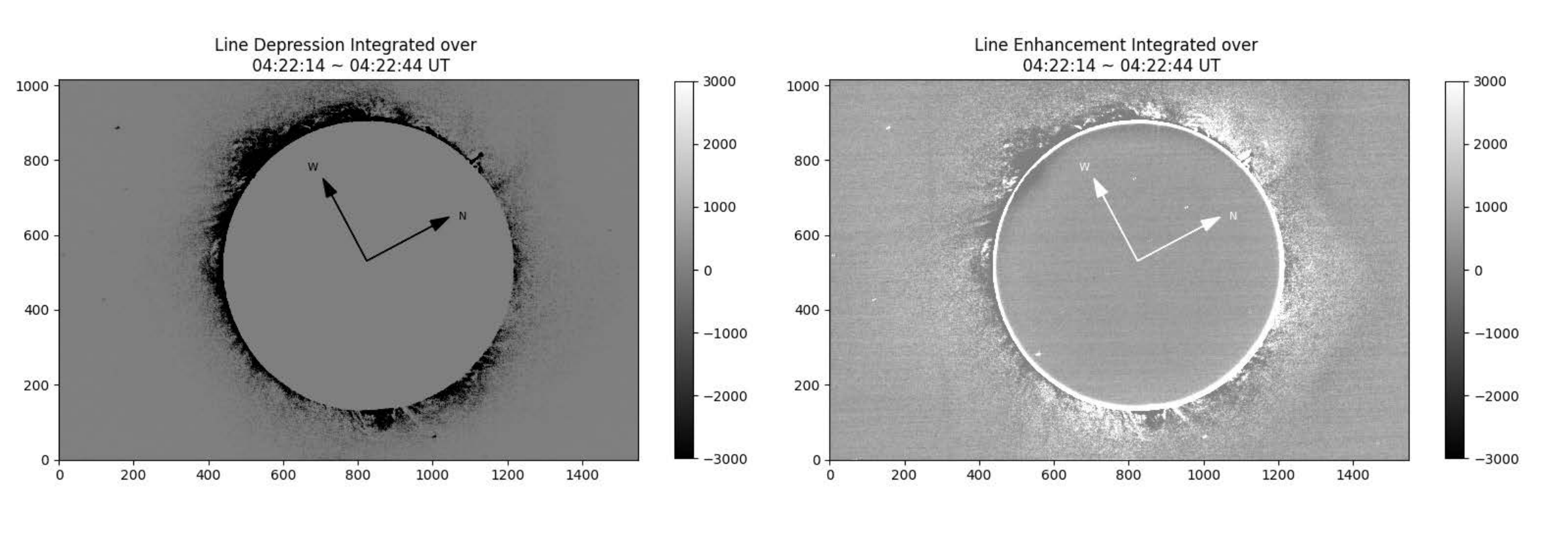}
\caption{\footnotesize Eclipse images. Top panels from left to
right: superposed maps obtained respectively from 659.4nm-centered
filtering observation and from 660.1nm-centered observation,
performed via the twin telescopes during the total solar eclipse on
20 Apr 2023. The 660.1nm-centered filtering image(on the top right)
represents the K-corona at this band. It is very hard to distinguish
these two images from each other on the top panels, since their
difference is very small. Middle panel: the map of intensity
difference between the two bands. Bottom panels: the map of the line
depression(left) and the map of the line enhancement(right). They
are respectively the maps of the inner F-corona and E-corona.}
\label{}
\end{figure}

The middle panel below the raw images is the map of the intensity
difference $I_{c,l}-I_{c}$ between the intensity of 659.4nm-centered
band $I_{c,l}$ and that of the 660.1nm-centered band $I_{c}$. All
the intensities are in unit of the detector readout counts(cts).
According to a simple calculation, the absolute value of the
normalized difference $(I_{c,l}-I_{c})/I_{c}$ is found to be less
than 0.3$\%$ in all these spatial points with intensities well above
the noise level. This means that the neutral iron atoms cannot be
easily observed at these two lines. As a contrast, we have found
that more than 20$\%$ line depression in many Fraunhofer lines
including those of the neutral iron, magnesium and chromium atoms in
the band 516.3nm-532.6nm was distorted in a very small coronal
region(Qu et al., 2024). And more popular is the several thousandth
line depression. Therefore, even the same kind of neutral atoms but
yielding different spectral lines can be distributed differently in
solar corona. This reflects different temperature distributions. The
weak correlation between the derived map and the K-corona one on the
top right means that the distribution of the neutral iron atoms is
very different from that of the free electrons. On the whole, the
line depression seems overwhelming, while the line enhancement is
more narrowly distributed. This reflects again the different
temperature distributions.

It is noteworthy that there was no detectable exotic lines found in
the two bands, as witnessed in Fig.2, where the spectra obtained
from the Sol$'$Ex portable spectrograph are presented. The spectrum
acquired before the second contact shows that the H$\alpha$
absorption line is the strongest and broadest in the band, while the
pair of neutral iron lines FeI659.4nm and FeI659.3nm are relatively
strong Fraunhofer lines compared to the other lines, and no spectral
lines can be detected within the band centered at 660.1nm with width
0.3nm. In the bottom panel is shown the spectrum acquired during
totality. Now, the  H$\alpha$ line intensity becomes prominent and
even saturate in those points close to the limb(the dark part
boarder), but the pair of the neutral iron lines now become so weak
that the 659.3nm line can be seen and the 659.4nm line can be
distinguished only by very carefully view. It is turned to be even
much weaker than other medium strong lines. This specifies that the
lines are not produced by the dust scattering, or the relative
intensities among these Fraunhofer lines should be kept. On the
other hand, there are no lines appeared within 0.3nm around 660.1nm.
All of these promise that the difference between intensities of
these two bands respectively centered at 659.4nm and 660.1nm is not
caused by other lines emerged during the totality.

The middle panel map is composed of the line depression(dark) and
the line enhancement(bright) simultaneously. It contains also those
noisy points with their intensity differences not well above the
noise level. In order to separate the line depression and the line
enhancement thus to get the F-corona and E-corona maps, we execute
an operation to divide the difference map into two. The F-corona map
depicted on the bottom left panel is obtained to collect those
points with their $(I_{c}-I_{c,l})>3\sigma$, and $\sigma$ indicates
the standard deviation evaluated at the solar disk occulted by the
moon during the totality, while the E-corona map shown in the bottom
right panel is attained if requirement $(I_{c,l}-I_{c})>3\sigma$ is
met.

It is impressive that the line depression can be strong in both the
quiet sun and active regions. Further, in the quiet sun region
around the south pole, the line depression is spread much more
widely than that in the domain around the prominence close to the
north pole, though the line depression can be found in the central
thin axis vertical to the local limb. In fact, the line depression
is much more concentrated in the active regions of the west
hemisphere, while the line enhancement locates more densely in the
hemisphere divided by a direction from the northwest to the
southeast. The diffusion motions of these neutral atoms can be seen
above both the active and quiet sun regions, indicated by the rapid
decrease of the line depression and enhancement with height above
the limb. Meanwhile, in the active region, the magnetic constraint
can indirectly influence the diffusion via collision of the neutral
atoms with these confined free electrons and ions. However, the line
depression can neither be viewed in the main body around the line
depression axis of the prominence, nor in the jet occurred close to
the east limb, and hardly observed in other domains dominated by the
line enhancement. Therefore, both of them are not homogeneously
spread around the solar disk occulted by the moon. Finally, it seems
that the line enhancement can detectable in higher regions, as
witnessed from the space above the prominence.

\begin{figure}
\vspace{3cm} \flushleft
\includegraphics[width=16.8cm,height=6.2cm]{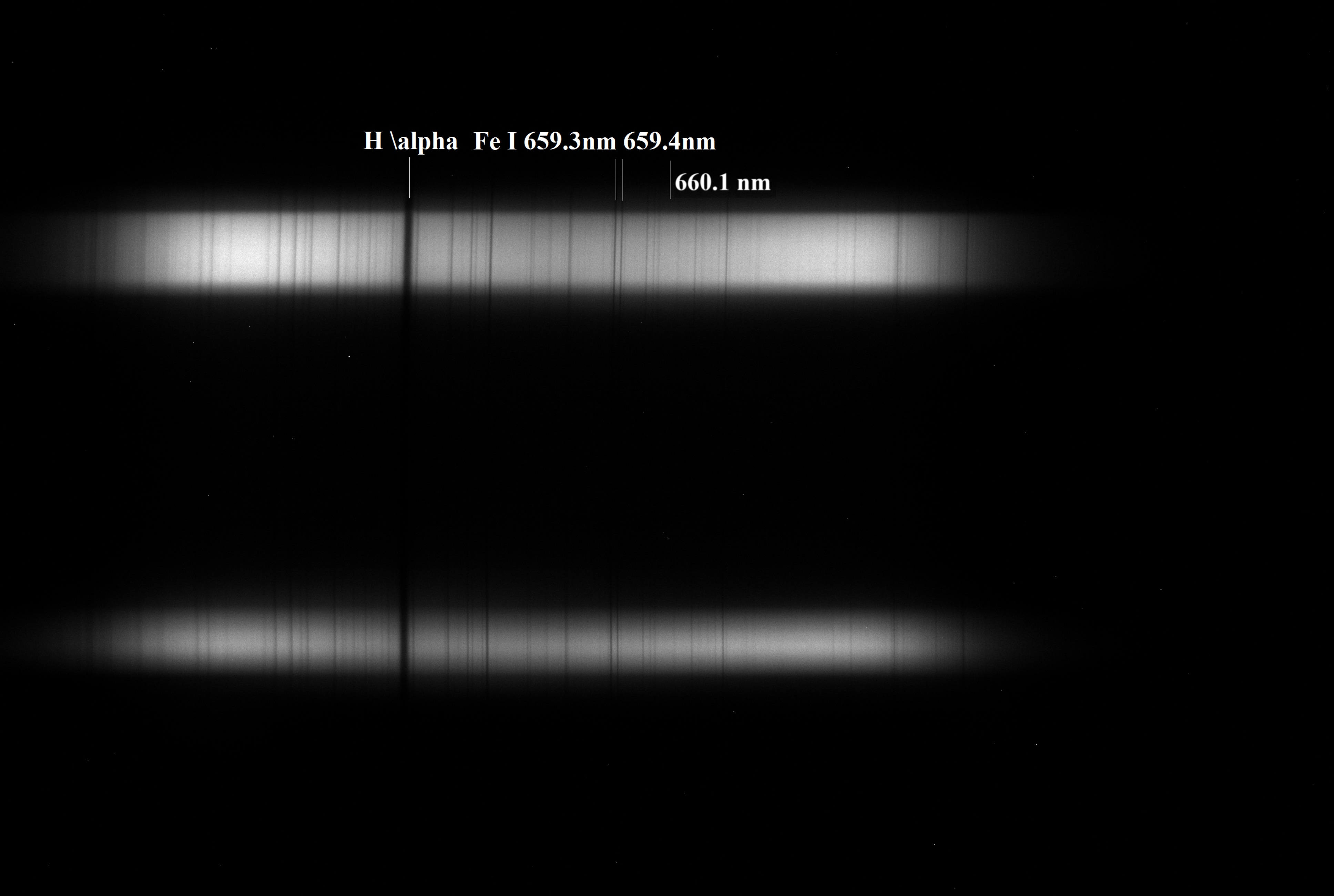}\\
\flushleft\vspace{0.2cm}
\includegraphics[width=16.8cm,height=6.2cm]{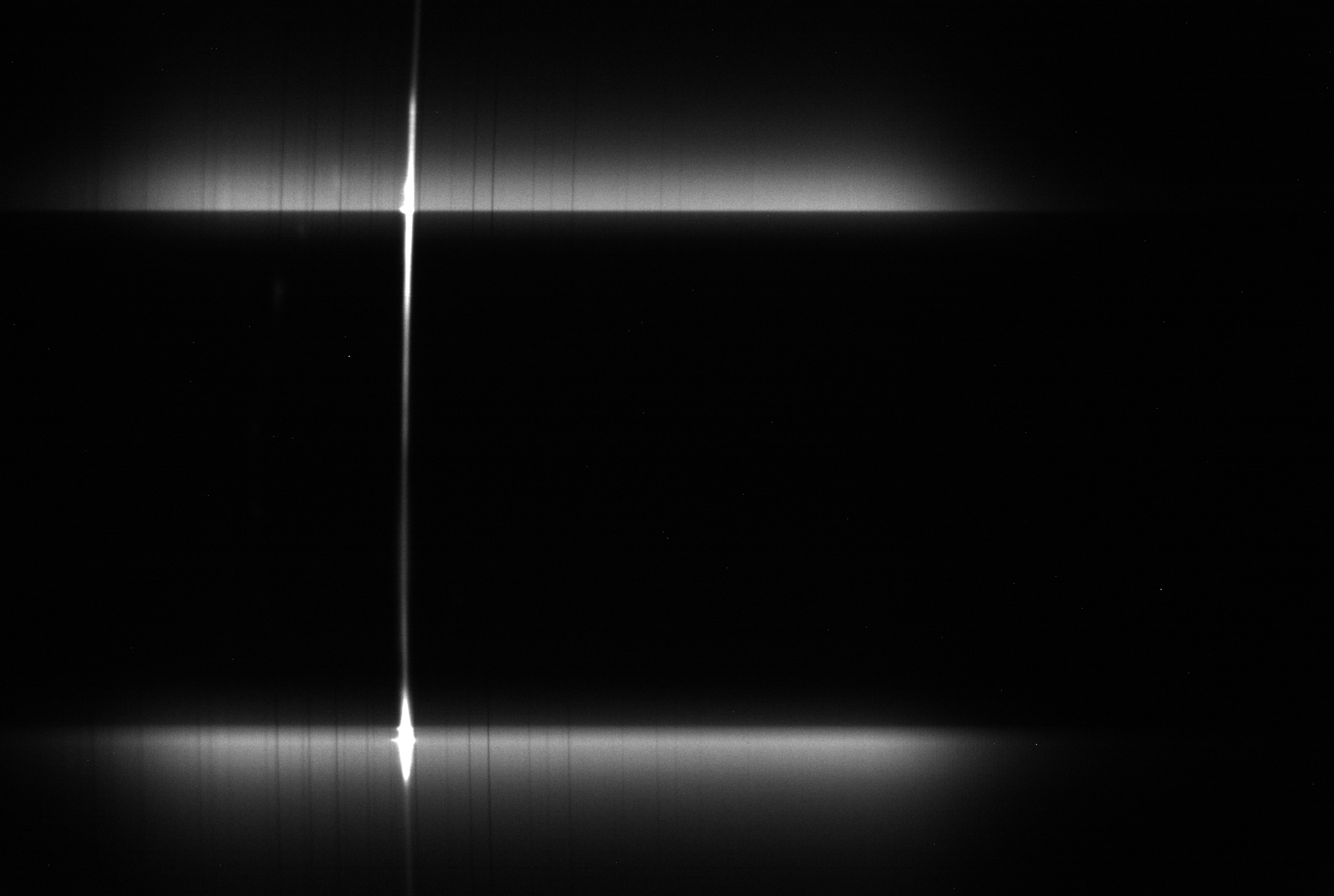}
\caption{\footnotesize Spectrum samples acquired from the portable
Sol$'$Ex spectrograph. Top panel: the spectrum acquired before the
second contact. The H$\alpha$ absorption line is the strongest and
broadest line in the band, while the pair of neutral iron lines
FeI659.4nm and FeI659.3nm are strong Fraunhofer lines among the
other lines, while no spectral lines can be seen within the band
centered at 660.1nm with width 0.3nm. Bottom panel: the spectrum
acquired during totality. The  H$\alpha$ line becomes dominant
emission one and saturated in spatial points close to the limb, but
the pair of the neutral iron lines become so weak that the 659.3nm
line can be hardly seen and the 659.4nm line can be visible by
careful checking in the lower part and turned to be much weaker than
other medium strong lines. This specifies the lines are not produced
by the dust scattering. On the other hand, there are no lines
appeared around 660.1nm within 0.3nm. This promises that the
difference between intensities of these two subbands respectively
centered at 659.4nm and 660.1nm is not caused by other lines emerged
during the totality.} \label{}
\end{figure}

The non-similarity between the line depression and enhancement maps
means the different formation temperature distribution of the
neutral iron atoms due to different excitation potentials of the
lower and upper energy levels involved in the transition.
Furthermore, it is easy to distinguish these two kinds of maps from
that of the K-corona depicted in the top right panel of Fig.1. It is
witnessed that though the K-coronal radiative intensity is much
stronger, it experiences much steeper decrease with the heliocentric
height than the E-corona or the F-corona.

It becomes clear that both the two kinds of distributions are
neither homogeneous nor symmetric. The F-corona map reflects the
distribution of the Fraunhofer line intensity depression,
proportional to net amount of photons taken away from the continuum
by particles staying at the lower levels of the involved
transitions, via scattering along the line-of-sight (Zanstra, 1941)
and/or absorption overwhelming emission. The E-corona map, on the
other hand, presents the line enhancement, proportional to net
amount of photons created by the atoms stayed at the upper energy
levels of the transitions above the continuum, via scattering and/or
emission overwhelming absorption. The K-corona map depicted on the
top right panel of Fig.1 provides quantitatively the distribution of
the free electrons predominantly responsible for the continuum
radiation. Therefore, till now, we can judge the presence of atoms
or ions responsible for the production of the specific spectral
lines within the inner corona concerned here. It is noteworthy that
the F-corona map induced by the neutral iron atoms presented here
looks very different from those caused from scattering by the dust
which exists above 3.0 solar radius height, distributed in an
elliptic space around the sun (Lamy et al., 2022; Burtovoi, 2022).
Furthermore, the line depression in these upper solar atmospheric
layers is evidently connected to those in the photosphere below,
because no line depression faultage is found in the chromosphere or
the transition zone.

\section{Conclusion}

After we obtain the F-corona and E-corona maps based on the line
depression and enhancement of the FeI659.3nm and 659.4nm lines, it
is judged that the neutral iron atoms are widely but sparsely
distributed in the inner solar corona as a cool component, with help
of the spectral images obtained by the portable Sol$'$Ex
spectrograph. The F-corona induced from the neutral iron atoms are
very different from that F-corona caused by dust scattering the
photospheric Fraunhofer lines, in aspects of components,
distribution patterns and heliocentric height ranges. Therefore, the
new F-corona forms the inner F-corona. The F-corona and E-corona
maps differ greatly from each other, and further both of them are
distinguished from K-corona critically. This reflects the different
distributions of particles with different temperatures.

The finding of the neutral atoms additional to the highly ionized
ions in the inner corona reveals a new nature of the solar corona.
The more micro atomic energy states of iron found leads to a
conclusion that the entropy estimated in the inner corona would be
much greater than that in the photosphere. Therefore, the coronal
heating can be understood according to the second law of the
thermodynamics. On the other hand, due to the presence of the
neutral atoms and thus ion-atom collision, the Cowling dissipation
will play a significant role in coronal heating(Cowling, 1957),
since the dissipation rate will become larger by many orders than
the rate of dissipation caused by electron-ion collisions. It also
provides a constraint on the coronal heating mechanism. That is, the
heating is not pervasive since a small fraction of neutral iron
atoms can escape from the heating into the corona, companying
upflows of other heated particles(Tian et al., 2021), through the
chromosphere and transition zone from the photosphere, where the
Fraunhofer lines are predominantly formed. Finally, the cool matter
injected into magnetic reconnection region can evidently promote the
energy release efficiencies(Ni et al., 2015; Liu et al., 2023;).
Therefore, the simulations of multi-fluid formed by the free
electrons, ions and neutral atoms sound more reasonable(Leake et
al., 2012) than the two-fluid model.

 \acknowledgments{\footnotesize This work is sponsored by
National Science Foundation of China (NSFC) under the grant numbers
11078005, U1931206 and 12003066.}

\end{document}